\pdfoutput=1
\documentclass{article}%
\usepackage{amssymb}
\usepackage{amsfonts}
\usepackage{amsmath}
\usepackage{graphicx}%
\providecommand{\U}[1]{\protect\rule{.1in}{.1in}}
\newtheorem{theorem}{Theorem}

\newtheorem{corollary}[theorem]{Corollary}

\newtheorem{lemma}[theorem]{Lemma}

\newtheorem{proposition}[theorem]{Proposition}

\begin{document}

\title{Time reversal in photoacoustic tomography and levitation in a cavity}
\author{V. P. Palamodov}
\date{}
\maketitle

School of Mathematical Sciences, Tel Aviv University, Tel Aviv, Israel

E-mail: palamodo@post.tau.ac.il

\bigskip

\textbf{Abstract }

A class of photoacoustic acquisition geometries in $\mathbb{R}^{n}$ is
considered such that the spherical mean transform admits an exact filtered
back projection reconstruction formula. The reconstruction is interpreted as a
time reversion mirror that reproduces exactly an arbitrary source distribution
in the cavity. A series of examples of non-uniqueness of the inverse potential
problem is constructed basing on the same geometrical technique.

\section{Introduction}

Reconstruction\ of a function from its spherical means is a mathematical tool
for photo- and thermo-acoustic tomography and SAR technic. Typically the
integral data is available for all spheres centered at a set $\mathrm{Z}%
\subset\mathbb{R}^{n}$ (e. g. an array of transducers) and an unknown function
is supported by an open set $\mathrm{H}\subset\mathbb{R}^{n}\backslash
\mathrm{Z}$. Closed reconstruction formulae of filtered back projection type
are known for few types of central sets$\ \mathrm{Z.}$ We state here that such
a reconstruction is possible for a class of algebraic central sets
$\mathrm{Z}$ called oscillatory \S \S 2-4. Next, we show that the
reconstructions can be interpreted as application of an universal time
reversal method for the wave equation$\mathrm{.}$ A signal is recorded by an
array of transducers (mirror), time-reversed and retransmitted into the
medium. The retransmitted signal propagates back through the same medium and
refocuses on the source. The time reversal method is an effective tool in
acoustical imaging even if the array of transducers is aside of a radiation
source and only a limit angle data are available (see \cite{Fink}). If arrays
completely surround the support of source distribution then complete angular
data can be detected.

We show that for a free space and any oscillatory array set $\mathrm{Z}$, the
time reversal is a perfect mirror providing the exact reconstruction of any
source distribution supported by the cavity $\mathrm{H}$ \S 5. The cavity can
be open; the method works for a paraboloid, a half-space and a two sheet
hyperboloid giving an approximate reconstruction with a finite array of
transducers. A strange point is that the time reversal method looking quite
natural in our problem works accurate only for very special class of center
sets $\mathrm{Z}$.

The method of oscillatory geometry can be applied also for a generalization of
famous Newton's attraction theorem (Principia, 1687). This theorem states that
a mass uniformly distributed over a thin sphere exerts zero gravitation field
inside (levitation). This is an example of the non-uniqueness of the interior
inverse potential problem. For a survey of inverse problems of potential
theory \cite{Z}. We show that there are many sets $\mathrm{Z}$ in
$\mathbb{R}^{3}$ supporting a mass distribution which generate levitation in
an open set \S \S 7-10.

\section{Oscillatory\ sets}

\textbf{Definition. }Let $p$ be a real polynomial in $\mathbb{R}^{n}$ of
degree $m$ with the zero set $\mathrm{Z}$. We call $p$ and $\mathrm{Z}$
\textit{oscillatory} with respect to a point $a\in\mathbb{R}^{n}%
\backslash\mathrm{Z},$ if $p$ has $m$ simple zeros in $L$ for almost any line
$L\subset\mathbb{R}^{n}$ through $a\mathrm{.}$

\begin{theorem}
\label{Con}\ Let $p$ be a polynomial in $\mathbb{R}^{n},$ $n>1$ with a compact
oscillatory zero set $\mathrm{Z}$ with respect to a point $a.$ We have
$\mathrm{Z}=\mathrm{Z}_{1}\cup...\cup\mathrm{Z}_{\mu},$ where $\mathrm{Z}%
_{1},...,\mathrm{Z}_{\mu},\ \mu=m/2$ are ovals (homeomorphic images of a $n-1$
sphere). They are nested in the sense that set of regular points of
$\mathrm{Z}_{i}$ is contained in the interior of $\mathrm{Z}_{i+1}\ $for
$i=1,...,\mu.$ Moreover, $\mathrm{Z}_{1}=\partial\mathrm{H}$ where
$\mathrm{H}$ is the set of all hyperbolic points. It is a convex component of
$\mathbb{R}^{n}\backslash\mathrm{Z.}$
\end{theorem}

\textbf{Proof. }Let $a$ be a hyperbolic point of $p$ and $\mathrm{S}$ be a
sphere in $\mathbb{R}^{n}$ with the center at the origin. For any$\ \omega
\in\mathrm{S}$ we numerate zero points of $p$ by $x_{k}\left(  \omega\right)
=a+t_{k}\omega,\ k=1,...,\mu\left(  \omega\right)  $ counting with
multiplicity in such a way that
\begin{equation}
t_{-\sigma\left(  \omega\right)  }\leq...\leq t_{-2}\leq t_{-1}<0<t_{1}\leq
t_{2}\leq...\leq t_{\tau\left(  \omega\right)  }. \label{16}%
\end{equation}
By Rouche's theorem for an arbitrary $\omega_{0}\in\mathrm{S}$ and arbitrary
$u<v,$ the number of zeros $t=t_{k}\left(  \omega\right)  $ of $p\left(
a+t\omega\right)  $ such that $u<t<v$ is constant for all $\omega$ in a
neighborhood of $\omega_{0}$ if $p\left(  a+u\omega_{0}\right)  p\left(
a+v\omega_{0}\right)  \neq0.$ Moreover, $t_{k}\left(  \omega\right)  $ is\ a
$1/m$-H\"{o}lder continuous function for any $k$. This implies $\sigma\left(
\omega\right)  +\tau\left(  \omega\right)  =m$ for any $\omega$ since
$\mathrm{Z}$ is compact. Because of the sphere $\mathrm{S}$ is connected, we
have $\sigma=\tau=\mu\doteqdot m/2$ and $t_{-k}\left(  \omega\right)
=t_{k}\left(  -\omega\right)  ,\ k=1,...,\mu$ for $\omega\in\mathrm{S}.$ For
any$\ k=1,...,\mu$, the function $x_{k}\left(  \omega\right)  =a+t_{k}\left(
\omega\right)  \omega$ is defined and continuous for $\omega\in\mathrm{S},$
hence$\ \mathrm{Z}_{k}=\left\{  x=x_{k}\left(  \omega\right)  ,\ \omega
\in\mathrm{S}\right\}  \ $is an oval. We have $\mathrm{Z}=\cup_{1}^{\mu
}\mathrm{Z}_{k}$ and the variety $\mathrm{Z}_{k}\cap\mathrm{Z}_{j}$ has
dimension $<n-1$ for any $j\neq k$. Therefore the sets $\mathrm{Z}_{k}$ are
nested and $a$ belongs to the interior $\mathrm{H}$ of the oval $\mathrm{Z}%
_{1}.$ It is easy to see that $\mathrm{Z}$ is oscillatory with respect to an
arbitrary $b\in\mathrm{H}$ and $\mathrm{H}$ is convex. We omit a detailed
proof which is geometrically transparent. $\blacktriangleright$

\textbf{Definition.\ }We call a \textit{hyperbolic cavity} of an oscillatory
set $\mathrm{Z}$ any maximal connected set $\mathrm{H}$ of points $a$ such
that $\mathrm{Z}$ is oscillatory with respect to $a$. By Theorem \ref{Con}
there is only one hyperbolic cavity, if the zero set is compact.

\textbf{Examples:\ 1. }Two sheet hyperboloid is oscillatory with two
hyperbolic\newline cavities, whereas any ellipsoid, elliptic paraboloid,
elliptic and parabolic cylinder is oscillatory with only one hyperbolic
cavity. A slab has three hyperbolic cavities. Other second order polynomials
are not oscillatory.

\textbf{2. }The zero set of the polynomial $p\doteq\left(  x^{2}+y^{2}\right)
^{3}-12\left(  x^{2}+y^{2}\right)  ^{2}+7x^{2}y^{2}+30\left(  x^{2}%
+y^{2}\right)  -20$ is compact, regular and oscillatory (see figure 1)%

\begin{center}
\fbox{\includegraphics[
natheight=3.000000in,
natwidth=4.499600in,
height=3in,
width=4.4996in
]%
{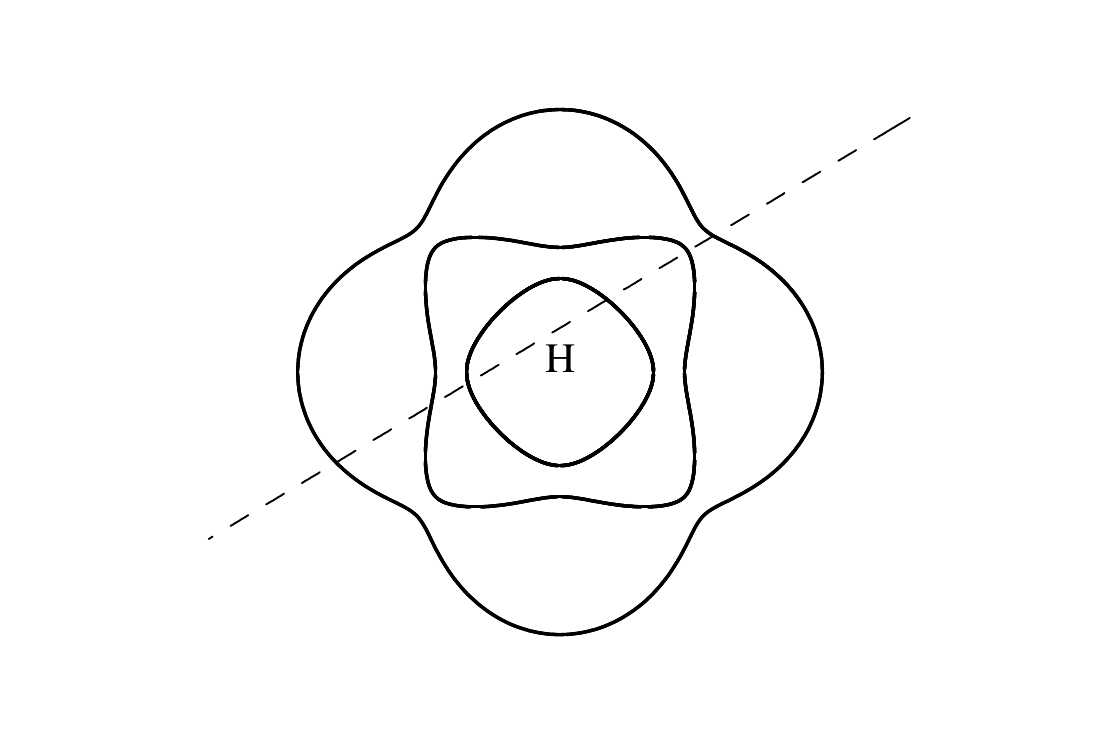}
}\\
Fig. 1 Oscillatory zero set of degree 6
\end{center}

\section{Photoacoustic inversion for oscillatory acquisition geometry}

Consider the spherical mean transform in an Euclidean space $E^{n}$ with a
central set $\mathrm{Z}\subset E^{n}$.
\[
\mathrm{R}_{\mathrm{Z}}f\left(  r,\xi\right)  )=\int_{\left\vert
x-\xi\right\vert =r}f\left(  x\right)  \mathrm{d}S,\ \xi\in\mathrm{Z},\ r>0
\]

\begin{theorem}
\label{H}Let $p$ be a polynomial in $E^{n}\ $with a compact regular
oscillatory zero set $\mathrm{Z}$ and $\mathrm{H}$ be the hyperbolic
cavity$\mathrm{.}$ An arbitrary\ function $f$ in $E^{n}$ with support in
$\mathrm{H}$ can be reconstructed from its spherical means by%
\begin{equation}
f\left(  x\right)  =-\frac{p\left(  x\right)  }{\mathrm{j}^{n-1}}%
\int_{\mathrm{Z}}\left(  \frac{1}{r}\frac{\partial}{\partial r}\right)
^{n-1}\left.  \frac{\mathrm{R}f\left(  r,\xi\right)  }{r}\right\vert
_{r=\left\vert x-\xi\right\vert }\frac{\mathrm{d}\xi}{\mathrm{d}p} \label{28}%
\end{equation}
for odd $n,$ and%
\begin{equation}
f\left(  x\right)  =-\frac{2p\left(  x\right)  }{\mathrm{j}^{n}}%
\int_{\mathrm{Z}}\int_{0}^{\infty}\frac{\mathrm{d}r^{2}}{\left\vert
x-\xi\right\vert ^{2}-r^{2}}\left(  \frac{1}{r}\frac{\partial}{\partial
r}\right)  ^{n-1}\frac{\mathrm{R}f\left(  r,\xi\right)  }{r}\frac
{\mathrm{d}\xi}{\mathrm{d}p} \label{9}%
\end{equation}
for even $n,$ where $\mathrm{Z}$ is oriented by the outward
conormal$\mathrm{.}$
\end{theorem}

A proof is given in \S 6.

Explicit reconstructions of filtered back projection type are due to Finch
\textit{et al} for the case $\mathrm{Z}$ is a sphere \cite{Fi2},\cite{Fi3}; Xu
and Wang \cite{XW} obtained explicit reconstruction when $\mathrm{Z}$ is a
sphere, a circular cylinder and a slab. Kunyanski \cite{K} considered some
polygonal curves and polyhedral surfaces as center sets. Recently a
reconstruction of FBP type $\mathrm{\ }$was done in \cite{Na} for ellipsoids
$\mathrm{Z}$ in 3D, and in \cite{Pa} for arbitrary dimension, see also
\cite{H}. The case of variable sound velocity is addressed in \cite{AK}.

\textbf{Example 3. }Half-space.\textbf{ }The central set $\mathrm{Z}=\left\{
x_{1}=0\right\}  $ in $E^{n},\ n\geq2$ is oscillatory with two hyperbolic
cavities$\mathrm{.}$ Such acquisition geometry appears in geophysics and in
the SAR technic. Reconstructions (\ref{28}) and (\ref{9}) for this case look
different from that of \cite{Fa} and \cite{An}. However, \textrm{Z }is not
compact and a regularization is necessary for application of either formula.

\textbf{4. }Half-ellipsoid\textbf{. }Let $p_{2}$ be a second order polynomial
with positive principle part. The set $p_{3}=0$ where $p_{3}=x_{1}p_{2}$ is
the union of an ellipsoid and a hyperplane. If the set $\mathrm{H}=\left\{
p_{2}\left(  x\right)  <0,x_{1}>0\right\}  $ is not empty it is a hyperbolic
cavity for $p_{3}$. We can approximate $p_{3}$ by a polynomial $\tilde{p}_{3}$
close to $p_{3}$with a regular oscillatory zero set $\mathrm{\tilde{Z}}$ (see
\S 9). The set $\mathrm{\tilde{Z}}$ consists of an oval $\mathrm{Z}%
_{1}=\partial\mathrm{H}$ and an unbounded component $\mathrm{Z}_{2}$ (see
figure 2):%
\begin{center}
\fbox{\includegraphics[
natheight=2.309100in,
natwidth=4.014500in,
height=2.3091in,
width=4.0145in
]%
{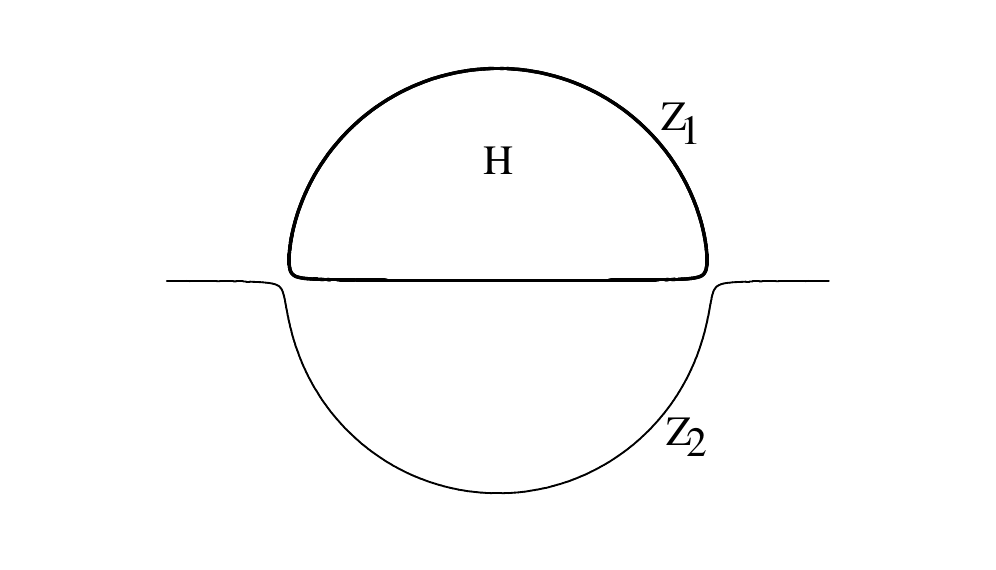}%
}\\
Fig. 2 Hyperbolic camera of a curve of degree 3
\end{center}
Note that contribution of points $\xi\in$ $\mathrm{Z}_{2}$ to (\ref{9})
decreases fast as $\xi\rightarrow\infty.$

\section{Time reversal structure}

Let $E^{n}$ be an Euclidean space; consider\ the Cauchy problem for the wave
equation in the space-time $\mathbb{R}\times E^{n}$%
\begin{align}
\left(  \frac{\partial^{2}}{\partial t^{2}}-\Delta\right)  u  &
=0\label{10}\\
u\left(  0,x\right)   &  =0,\ u_{t}^{\prime}\left(  0,x\right)  =f\left(
x\right)  .\nonumber
\end{align}
for a function $f$ in $E^{n}.$ Let $E_{n+1}\left(  t,x\right)  $ be the
forward propagator for (\ref{10}).

\begin{theorem}
Formulae (\ref{28}) and (\ref{9}) are equivalent to the time reversal method
for (\ref{10}) acting in the following steps:\newline(i) transmission (forward
propagation) of a function $f$ supported in $\mathrm{H}$ to the mirror
manifold$\ \mathbb{R}\times\mathrm{Z}$,%
\[
f\mapsto u\left(  t,\xi\right)  =\int_{\mathrm{H}}E_{n+1}\left(
t,x-\xi\right)  f\left(  x\right)  \mathrm{d}x,\ \xi\in\mathrm{Z,}%
\]
\newline(ii) filtration%
\[
vt=\mathrm{F}u\doteqdot-\frac{\partial}{\partial t}\frac{2}{t}\frac{\partial
}{\partial t}u,
\]
\newline(iii) time reversion and retransmission
\[
g\left(  x\right)  =\int_{\mathrm{Z}}\int_{0}^{\infty}E_{n+1}\left(
t,x-\xi\right)  v\left(  -t,\xi\right)  \mathrm{d}t\frac{\mathrm{d}\xi
}{\mathrm{d}p},
\]
(iv) reconstruction$\ $%
\[
f\left(  x\right)  =-2p\left(  x\right)  g\left(  x\right)  .
\]%
\end{theorem}

\begin{center}
\fbox{\includegraphics[
natheight=3.000000in,
natwidth=4.498700in,
height=3in,
width=4.4987in
]%
{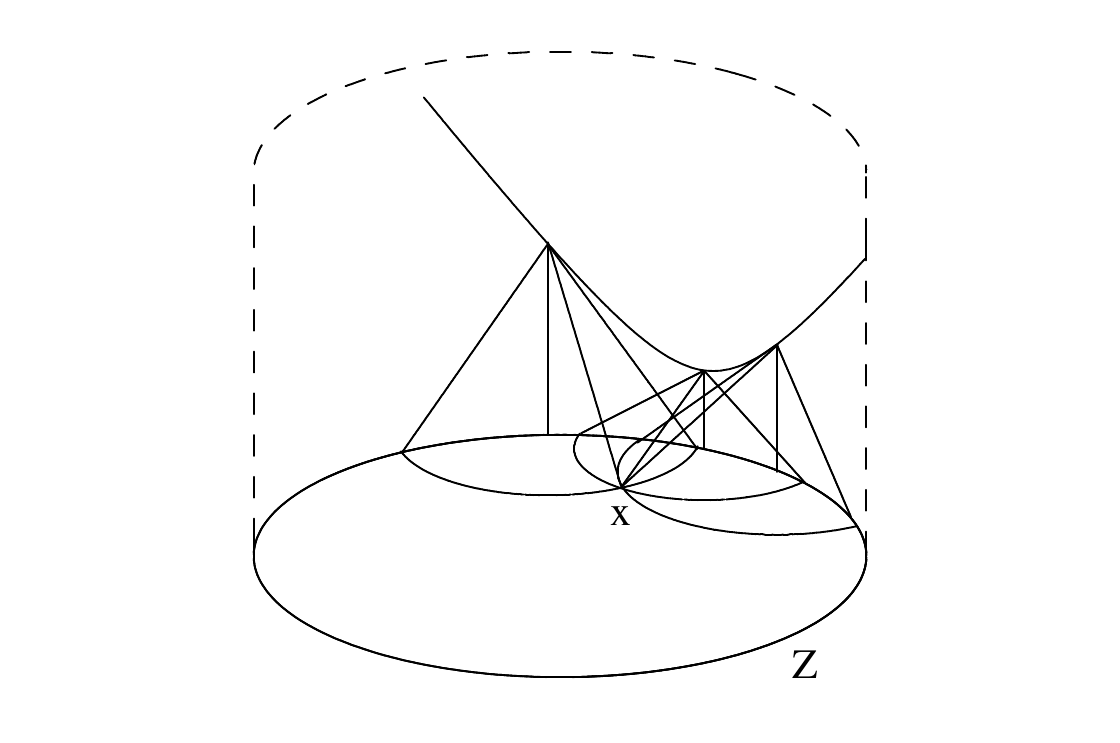}%
}\\
Fig. 3 Geometry of the time reversal
\end{center}

\textbf{Remark.} The filtration operator $\mathrm{F}$ is a positive
self-adjoint differential operator in the Hilbert space $L_{2}\left(
\mathbb{R}_{+}\right)  .$ We can replace the volume form $\mathrm{d}%
\xi/\mathrm{d}p$ by $q\mathrm{d}\xi/\mathrm{d}p$ where $q$ is an arbitrary
polynomial of degree $\leq m-2$ preserving equations (\ref{28}) and (\ref{9}).
If $q$ is a strict separator of $\mathrm{Z}$ (\S 8), the form $q\mathrm{d}%
x/\mathrm{d}p$ is a volume form in $\mathrm{Z.}$ Then the retransmitting
operator $\mathrm{E}^{\ast}$ is adjoint to the transmitting operator
$\mathrm{E}:$ $L_{2}\left(  \mathrm{H}\right)  \rightarrow L_{2}\left(
\mathbb{R}\times\mathrm{Z}\right)  $ where the space $\mathbb{R}%
\times\mathrm{Z}$ is endowed with the volume form $q\mathrm{d}x/\mathrm{d}p.$
The time reversal operator $\mathrm{T}$ can be written in the self-adjont form%
\[
\mathrm{T}=\mathrm{E}^{\ast}\mathrm{FE.}%
\]

\textbf{Proof. }The forward propagators are%
\begin{align*}
E_{n+1}\left(  t,x\right)   &  =\frac{1}{2\pi}\left(  \frac{\partial}%
{\pi\partial t^{2}}\right)  ^{\left(  n-3\right)  /2}\theta\left(  t\right)
\delta\left(  t^{2}-\left\vert x\right\vert ^{2}\right)  \ \text{\ for odd
}n,\\
E_{n+1}\left(  t,x\right)   &  =\frac{1}{2\pi}\left(  \frac{\partial}%
{\pi\partial t^{2}}\right)  ^{\left(  n-2\right)  /2}\frac{\theta\left(
t-\left\vert x\right\vert \right)  }{\left(  t^{2}-\left\vert x\right\vert
^{2}\right)  ^{1/2}}\ \ \text{for even }n.
\end{align*}

Let $n$ be \textbf{odd.} The solution of (\ref{10}) equals%
\[
u\left(  t,\xi\right)  =\frac{1}{4\pi}\left(  \frac{\partial}{\pi\partial
t^{2}}\right)  ^{\left(  n-3\right)  /2}\frac{1}{t}\int_{\left\vert
x-\xi\right\vert =t}f\mathrm{d}S=\frac{1}{4\pi}\left(  \frac{\partial}%
{\pi\partial t^{2}}\right)  ^{\left(  n-3\right)  /2}\frac{\mathrm{R}f\left(
t,\xi\right)  }{t}%
\]
and by (\ref{28})%
\begin{align*}
-\frac{f\left(  x\right)  }{2p\left(  x\right)  }  &  =\frac{2^{n-2}%
}{\mathrm{j}^{n-1}}\int_{\mathrm{Z}}\frac{\mathrm{d}\xi}{\mathrm{d}p}\left(
\frac{\partial}{\partial t^{2}}\right)  ^{n-1}\left.  \frac{\mathrm{R}f\left(
t,\xi\right)  }{t}\right\vert _{t=\left\vert x-\xi\right\vert }\\
&  =\left(  -1\right)  ^{\left(  n-1\right)  /2}\frac{1}{2\pi}\int
_{\mathrm{Z}}\frac{\mathrm{d}\xi}{\mathrm{d}p}\left(  \frac{\partial}%
{\pi\partial t^{2}}\right)  ^{\left(  n-3\right)  /2}\left.  \frac{1}{t}%
\frac{\partial}{\partial t}\frac{1}{t}\frac{\partial}{\partial t}u\left(
t,\xi\right)  \right\vert _{t=\left\vert x-\xi\right\vert }\\
&  =-\frac{1}{\pi}\int_{\mathrm{Z}}\int_{0}^{\infty}\left(  \frac{\partial
}{\pi\partial t^{2}}\right)  ^{\left(  n-3\right)  /2}\delta\left(
t^{2}-\left\vert x-\xi\right\vert ^{2}\right)  \frac{\partial}{\partial
t}\frac{1}{t}\frac{\partial}{\partial t}u\left(  t,\xi\right)  \mathrm{d}%
t\frac{\mathrm{d}\xi}{\mathrm{d}p}\\
&  =\int_{\mathrm{Z}}\int_{\mathbb{R}}\left(  E_{n+1}\left(  t,x-\xi\right)
\ast v\left(  -t,\xi\right)  \right)  \mathrm{d}t\frac{\mathrm{d}\xi
}{\mathrm{d}p}\\
&  =\int_{\mathrm{Z}}\int_{\mathbb{R}}\left(  E_{n+1}\left(  -t,x-\xi\right)
\ast v\left(  t,\xi\right)  \right)  \mathrm{d}t\frac{\mathrm{d}\xi
}{\mathrm{d}p}%
\end{align*}
is the retransmission of $u,$ where $v=\mathrm{F}u$.

For \textbf{even }$n$ we consider only the case $n=2.$

\begin{lemma}
\label{HA}We have%
\[
\int_{-\infty}^{\infty}\frac{\mathrm{d}\rho}{\left(  \sigma-\rho\right)
\left(  \rho-\tau\right)  _{+}^{1/2}}=\frac{\pi}{\left(  \tau-\sigma\right)
_{+}^{1/2}}%
\]
where $t_{\pm}\doteq\max\left\{  \pm t,0\right\}  .$
\end{lemma}

A proof can be done by application of the Fourier transform.
$\blacktriangleright$

By (\ref{9}) for $n=2,$
\[
-\frac{f\left(  x\right)  }{2p\left(  x\right)  }=\frac{1}{\mathrm{j}^{2}}%
\int_{\mathrm{Z}}\frac{\mathrm{d}\xi}{\mathrm{d}p}\int_{0}^{\infty}%
\frac{\mathrm{d}r^{2}}{\left\vert x-\xi\right\vert ^{2}-r^{2}}\frac{1}{r}%
\frac{\partial}{\partial r}\frac{\mathrm{R}f\left(  r,\xi\right)  }{r}.
\]
Direct propagation by Poisson's formula%
\begin{align*}
u\left(  t,\xi\right)   &  =\frac{1}{2\pi}\int_{\left\vert x-\xi\right\vert
\leq t}\frac{f\left(  x\right)  \mathrm{d}x}{\left(  t^{2}-\left\vert
x-\xi\right\vert ^{2}\right)  ^{1/2}}=\frac{1}{4\pi}\int_{0}^{\tau}%
\frac{\mathrm{d}\sigma}{\left(  \tau-r^{2}\right)  ^{1/2}}\frac{\mathrm{R}%
f\left(  r,\xi\right)  }{r}\\
&  =\frac{1}{4\pi}\int_{-\infty}^{\infty}\frac{\mathrm{d}\sigma}{\left(
\tau-\sigma\right)  _{+}^{1/2}}\mathrm{S}f\left(  \sigma,\xi\right)  ,
\end{align*}
where$\ \ \tau=t^{2},\ r=\left\vert x-\xi\right\vert ^{2}$, $\mathrm{S}%
f\left(  \sigma,\xi\right)  =r^{-1}\mathrm{R}f\left(  r,\xi\right)  $ for
$\sigma=r^{2}$ and $\mathrm{S}f\left(  \sigma,\xi\right)  =0$ for $\sigma<0. $
Solving by Abel's method we find%
\[
\mathrm{S}f\left(  \sigma,\xi\right)  =4\frac{\mathrm{d}}{\mathrm{d}\rho}%
\int_{-\infty}^{\sigma}\frac{u\left(  t,\xi\right)  }{\left(  \sigma
-\tau\right)  ^{1/2}}\mathrm{d}\tau,
\]
where $u\left(  t,\xi\right)  =0$ for $\tau<0.$ We set $\rho=r^{2}$ in
(\ref{9}) and by Lemma \ref{HA} below%
\begin{align*}
-\frac{f\left(  x\right)  }{2p\left(  x\right)  }  &  =-\frac{1}{2\pi^{2}}%
\int_{\mathrm{Z}}\frac{\mathrm{d}\xi}{\mathrm{d}p}\int_{\mathbb{R}}%
\frac{\mathrm{d}\rho}{\sigma-\rho}\frac{\partial}{\partial\rho}\mathrm{S}%
f\left(  \rho,\xi\right) \\
\int_{\mathbb{R}}\frac{\mathrm{d}\rho}{\sigma-\rho}\frac{\partial}%
{\partial\rho}\mathrm{S}f\left(  \rho,\xi\right)   &  =\int_{\mathbb{R}%
}u\left(  t,\xi\right)  \mathrm{d}\tau\frac{\mathrm{d}\rho}{\sigma-\rho
}\left(  \frac{\partial}{\partial\rho}\right)  ^{2}\frac{1}{\left(  \rho
-\tau\right)  _{+}^{1/2}}\\
&  =\int_{\mathbb{R}}\left(  \frac{\partial}{\partial\sigma}\right)  ^{2}%
\frac{u\left(  t,\xi\right)  \mathrm{d}\tau}{\left(  \tau-\sigma\right)
_{+}^{1/2}}=\int_{\mathbb{R}}\left(  \frac{\partial}{\partial\tau}\right)
^{2}u\left(  t,\xi\right)  \frac{\mathrm{d}\tau}{\left(  \tau-\sigma\right)
_{+}^{1/2}}\\
&  =\int_{\mathbb{R}}\frac{\partial}{\partial t}\frac{2}{t}\frac{\partial
}{\partial t}u\left(  t,\xi\right)  \frac{\mathrm{d}t}{\left(  \tau
-\sigma\right)  _{+}^{1/2}}=-\int_{\mathbb{R}}\frac{v\left(  t,\xi\right)
\mathrm{d}t}{\left(  t^{2}-\left\vert x-\xi\right\vert ^{2}\right)  ^{1/2}}.
\end{align*}
since$\ \tau=t^{2},$ $\mathrm{d}\tau=2t\mathrm{d}t,\ \partial/\partial
\tau=1/2t\ \partial/\partial t$.\ This yields%
\begin{align*}
-\frac{f\left(  x\right)  }{2p\left(  x\right)  }  &  =\frac{1}{2\pi}%
\int_{\mathrm{Z}}\frac{\mathrm{d}\xi}{\mathrm{d}p}\int_{\mathbb{R}}%
\frac{v\left(  t,\xi\right)  \mathrm{d}t}{\left(  t^{2}-\left\vert
x-\xi\right\vert ^{2}\right)  ^{1/2}}\\
&  =\int_{\mathrm{Z}}E_{3}\left(  -t,x-\xi\right)  v\left(  t,\xi\right)
\frac{\mathrm{d}\xi}{\mathrm{d}p},
\end{align*}
which completes the proof. $\blacktriangleright$

\section{Proof of the reconstruction}

\textbf{Proof. }Check that the generating function $\Phi\left(  x;\lambda
,\xi\right)  =\theta\left(  x,\xi\right)  -\lambda,\ \theta=\left\vert
x-\xi\right\vert ^{2}$ defined in $\mathrm{H}\times\Sigma,\ \Sigma
=\mathbb{R}\times\mathrm{Z}$ satisfies conditions of Theorem 3.1 of
\cite{Pa}\textbf{.} Condition (\textbf{i) }is easy to check. To prove
(\textbf{ii}) we\textbf{\ }suppose that $y\neq x$ are conjugate points in
$\mathrm{H}$, which means%
\begin{equation}
\theta\left(  x,\xi\right)  =\theta\left(  y,\xi\right)  ,\ \mathrm{d}_{\xi
}\theta\left(  x,\xi\right)  =\mathrm{d}_{\xi}\theta\left(  y,\xi\right)
\label{22}%
\end{equation}
for some $\xi\in\mathrm{Z.}$ The first equation (\ref{22}) implies that
$\left\vert x-\xi\right\vert =\left\vert y-\xi\right\vert .$ It follows that
the line $L$ through $s=1/2\left(  x+y\right)  $ and $\xi$ is orthogonal to
$x-y.$ By the second condition (\ref{22}) we have $\left\langle x-y,\mathrm{d}%
\xi\right\rangle ,$ hence vector $x-y$ is orthogonal to $\mathrm{Z}$ at $\xi.$
Therefore $L$ is tangent to $\mathrm{Z}$ at $\xi$ which is impossible, since
$\mathrm{H}$ is convex and no line through $s\in\mathrm{H} $ is tangent to
$\mathrm{Z}$.\ Consider an integral
\[
\Theta_{n}\left(  x,y\right)  =\int_{\mathrm{Z}}\frac{\mathrm{\omega
}_{\mathrm{Z}}}{\left(  \varphi\left(  x,y;\xi\right)  -i0\right)  ^{n}},
\]
where%
\[
\varphi\left(  x,y;\xi\right)  =\theta\left(  y,\xi\right)  -\theta\left(
x,\xi\right)  =2\left\langle x-y,\xi\right\rangle +\left\vert y\right\vert
^{2}-\left\vert x\right\vert ^{2}=\left\langle z,\xi-s\right\rangle
,\ z=2\left(  x-y\right)  .
\]

\begin{lemma}
\label{3}We have $\operatorname{Re}i^{n}\Theta_{n}\left(  x,y\right)  =0$ for
arbitrary $y\neq x$.
\end{lemma}

\textbf{Proof. }Let $\mathrm{S}$ be the unit sphere in $E^{n}.$ We can write
$\mathrm{d}p=p_{t}^{\prime}\mathrm{d}t+\mathrm{d}_{\omega}p$ by means of
spherical coordinates $\xi=a+t\omega,\ t\in\mathbb{R},\ \omega\in\mathrm{S.}$
This yields
\begin{equation}
\mathrm{d}\xi=t^{n-1}\mathrm{d}t\wedge\mathrm{\Omega}=t^{n-1}\frac
{\mathrm{d}p}{p_{t}^{\prime}}\wedge\mathrm{\Omega},\ \frac{\mathrm{d}\xi
}{\mathrm{d}p}=\frac{t^{n-1}}{p_{t}^{\prime}}\mathrm{\Omega}, \label{5}%
\end{equation}
\textbf{\ }where $\mathrm{\Omega}$ is the volume form in the sphere
$\mathrm{S}$. We move the origin to the point $s\in\mathrm{H}$ and have
$\varphi\left(  \xi\right)  =\left\langle z,\xi\right\rangle .$ Let $t_{-\mu
}<..<t_{-1}<0<t_{1}<...<t_{\mu}$ be all zeros of $p\left(  a+t\omega\right)  $
as in Theorem \ref{Con}.\ For \textbf{even} $n,$ we have
\begin{equation}
2\operatorname{Re}\Theta_{n}\left(  x,y\right)  =2\int_{\mathrm{Z}}\frac
{1}{\varphi^{n}}\frac{\mathrm{d}\xi}{\mathrm{d}p}=\lim_{\varepsilon
\rightarrow0+}\int_{\mathrm{Z}}\left[  \frac{1}{\varphi\left(  t\omega
_{+}\right)  ^{n}}+\frac{1}{\varphi\left(  t\omega_{-}\right)  ^{n}}\right]
\frac{\mathrm{d}\xi}{\mathrm{d}p}, \label{6}%
\end{equation}
where $\omega_{\pm}=\omega\pm i\varepsilon\left\vert z\right\vert ^{-2}z, $
$\varepsilon>0$ is a small number. We have $\varphi\left(  t\omega_{+}\right)
=\left\langle tz,\omega_{\pm}\right\rangle =\left\langle tz,\omega
\right\rangle \pm i\varepsilon\mathrm{sgn}t$ and%
\[
\frac{1}{\varphi\left(  t\omega_{+}\right)  ^{n}}+\frac{1}{\varphi\left(
t\omega_{-}\right)  ^{n}}=\frac{1}{t^{n}}\left(  \frac{1}{\left\langle
z,\omega_{+}\right\rangle ^{n}}+\frac{1}{\left\langle z,\omega_{-}%
\right\rangle ^{n}}\right)
\]
for any $t\neq0.$ Taking in account (\ref{5}), we integrate over $\mathrm{Z} $
and get%
\begin{equation}
2\operatorname{Re}\Theta_{n}\left(  x,y\right)  =\lim_{\varepsilon
\rightarrow0}\int_{\mathrm{S}_{+}}\sum_{k=-\mu}^{\mu}\frac{1}{t_{k}%
p_{t}^{\prime}\left(  a+t_{k}\omega\right)  }\left[  \frac{1}{\left\langle
z,\omega_{+}\right\rangle ^{n}}+\frac{1}{\left\langle z,\omega_{-}%
\right\rangle ^{n}}\right]  \mathrm{\Omega}, \label{18}%
\end{equation}
where $\mathrm{S}_{+}$ is an arbitrary hemisphere. The sum in (\ref{18}) is
equal to the sum of residues $\mathrm{res}_{t_{k}}\rho\left(  t,\omega\right)
$ of the form$\ \rho\left(  t,\omega\right)  =$ $\mathrm{d}t/p\left(
a+t\omega\right)  t.$ Integrate this form$\ $along a circle of radius
$R>\max_{\mathrm{S}}\left\vert t_{\mu}\left(  \omega\right)  \right\vert $ and
apply the Residue theorem:%
\begin{align*}
\frac{1}{2\pi i}\int_{\left\vert t\right\vert =R}\rho\left(  t,\omega\right)
&  =\mathrm{res}_{0}\rho\left(  t,\omega\right)  +\sum_{k=-\mu}^{\mu
}\mathrm{res}_{t_{k}}\rho\left(  t,\omega\right)  =\frac{1}{p\left(  a\right)
}+\sum_{k=-\mu}^{\mu}\frac{1}{t_{k}p_{t}^{\prime}\left(  t_{k}\omega\right)
}\\
&  =-\mathrm{res}_{\infty}\rho\left(  t,\omega\right)  =0.
\end{align*}
Here$\ t_{k}\omega\in\mathrm{Z}_{\left\vert k\right\vert },\ k=-\mu,...,\mu$
and the residue at infinity vanishes since $\rho\left(  t,\omega\right)
=O\left(  t^{-2}\right)  .$ Therefore
\begin{equation}
\sum_{k=-\mu}^{\mu}\frac{1}{t_{k}p_{t}^{\prime}\left(  a+t_{k}\omega\right)
}=-\frac{1}{p\left(  a\right)  } \label{12}%
\end{equation}
and we come up with the equation%
\begin{align*}
2\operatorname{Re}\Theta_{n}\left(  x,y\right)   &  =-\frac{1}{p\left(
a\right)  }\lim_{\varepsilon\rightarrow0}\int_{\mathrm{S}_{+}}\left(  \frac
{1}{\left\langle z,\omega_{+}\right\rangle ^{n}}+\frac{1}{\left\langle
z,\omega_{-}\right\rangle ^{n}}\right)  \mathrm{\Omega}\\
&  =-\frac{1}{2p\left(  a\right)  }\operatorname{Re}\int_{\mathrm{S}}%
\frac{\mathrm{\Omega}}{\left(  \left\langle z,\omega\right\rangle -i0\right)
^{n}}.
\end{align*}
The right hand side vanishes by \cite{Pa} Proposition 4.3 hence
$\operatorname{Re}\Theta_{n}\left(  x,y\right)  =0.$

For \textbf{odd }$n,$ we argue in the similar way:%
\begin{align}
2i\operatorname{Im}\Theta_{n}\left(  x,y\right)   &  =\lim_{\varepsilon
\rightarrow0}\sum_{k=-\mu}^{\mu}\int_{\mathrm{S}_{+}}\frac{1}{t_{k}p^{\prime
}\left(  a+t_{k}\omega\right)  }\left[  \frac{1}{\left\langle z,\omega
_{-}\right\rangle ^{n}}-\frac{1}{\left\langle z,\omega_{+}\right\rangle ^{n}%
}\right]  \mathrm{\Omega}\nonumber\\
&  =-\frac{2i}{p\left(  a\right)  }\operatorname{Im}\int_{\mathrm{S}^{n-1}%
}\frac{\mathrm{\Omega}}{\left(  \left\langle z,\omega\right\rangle -i0\right)
^{n}}. \label{7}%
\end{align}
The right hand side vanishes according to \cite{Pa} Proposition 4.3. This
together with (\ref{7}) implies vanishing of $\operatorname{Im}\Theta
_{n}\left(  x,y\right)  $ and completes the proof of Lemma \ref{3}.
$\blacktriangleright$

Theorem \ref{H} now follows from \cite{Pa} Theorem 3.1\ applied for the
generating function $\Phi$ as above and for the space $\Sigma=\mathbb{R}%
\times\mathrm{Z}$ endowed with the form $\mathrm{d}x/\mathrm{d}p.$ We change
the variable $\lambda=r^{2}\ $and take into account that $\left\vert
\nabla\theta\right\vert =2\left\vert x-\xi\right\vert =2r$ and $\mathrm{M}%
f\left(  r^{2},x\right)  =\left(  2r\right)  ^{-1}\mathrm{R}f\left(
r,x\right)  \ $in loc. cit. To complete the proof we only need to calculate
the dominator
\[
D_{n}\left(  x\right)  =\frac{1}{\left\vert \mathrm{S}^{n-1}\right\vert }%
\int_{\mathrm{Z}}\frac{1}{\left\vert \xi-x\right\vert }\frac{\mathrm{d}\xi
}{\mathrm{d}p\left(  \xi\right)  }%
\]
for an arbitrary $x\in\mathrm{H.}$ For any $\xi\in\mathrm{Z},$ we can write
$\xi=x+t_{k}\left(  \omega\right)  \omega\ $for a unique $\omega\in\mathrm{S}$
and $t_{k}>0$ and have $\left\vert \xi-x\right\vert =t_{k}\left(
\omega\right)  .$ Replacing $a$ by $x$ in (\ref{5}) yields
\[
D_{n}\left(  x\right)  =\frac{1}{\left\vert \mathrm{S}^{n-1}\right\vert }%
\int_{\mathrm{S}}\sum_{k=1}^{\mu}\frac{\mathrm{\Omega}}{t_{k}p_{t}^{\prime
}\left(  x+t_{k}\left(  \omega\right)  \omega\right)  }.
\]
The sum of contributions of opposite points $\omega\in\mathrm{S}_{+}$ and
$-\omega$ equals%
\begin{align*}
&  \sum_{k=1}^{\mu}\frac{\mathrm{\Omega}}{t_{k}\left(  \omega\right)
p_{t}^{\prime}\left(  x+t_{k}\left(  \omega\right)  \omega\right)  }%
-\sum_{k=1}^{\mu}\frac{\mathrm{\Omega}}{t_{k}\left(  -\omega\right)
p_{s}^{\prime}\left(  x-t_{k}\left(  -\omega\right)  \omega\right)  }\\
&  =\sum_{k=-\mu}^{\mu}\frac{\mathrm{\Omega}}{t_{k}\left(  \omega\right)
p_{t}^{\prime}\left(  x+t_{k}\omega\right)  },
\end{align*}
where $s=-t$ and $t_{k}\left(  -\omega\right)  =-t_{-k}\left(  \omega\right)
,\ k=1,...,\mu.$ By (\ref{12}) this sum is equal to $-1/p\left(  x\right)  ,$
hence%
\[
D_{n}\left(  x\right)  =-\frac{1}{\left\vert \mathrm{S}^{n-1}\right\vert
p\left(  x\right)  }\int_{\mathrm{S}_{+}}\mathrm{\Omega}=-\frac{1}{2p\left(
x\right)  },
\]
which completes the proof of (\ref{28}) and (\ref{9}).$\ \blacktriangleright$

\section{Separators}

\textbf{Definition. }Let $p$ be an oscillatory polynomial of degree $m>2$ with
a hyperbolic point $a$. We say that a polynomial $q$ \textit{separates} $p$
from $a\mathrm{,}$ if for almost any line $L$ through $a,$ each
interval\ between consecutive zeros of $p$ on $L$ contains just one zero of
$q$, except for the interval that contains $a$, where $q$ does not vanish (see
figure 4). It follows that $q$ has, at least, $m-2$ zero on $L$, hence $\deg
q\geq m-2.$ We say that a separator $q$ of a polynomial $p$ is \textit{strict}
if $\deg q=m-2.$%

\begin{center}
\fbox{\includegraphics[
natheight=3.000000in,
natwidth=4.499600in,
height=3in,
width=4.4996in
]%
{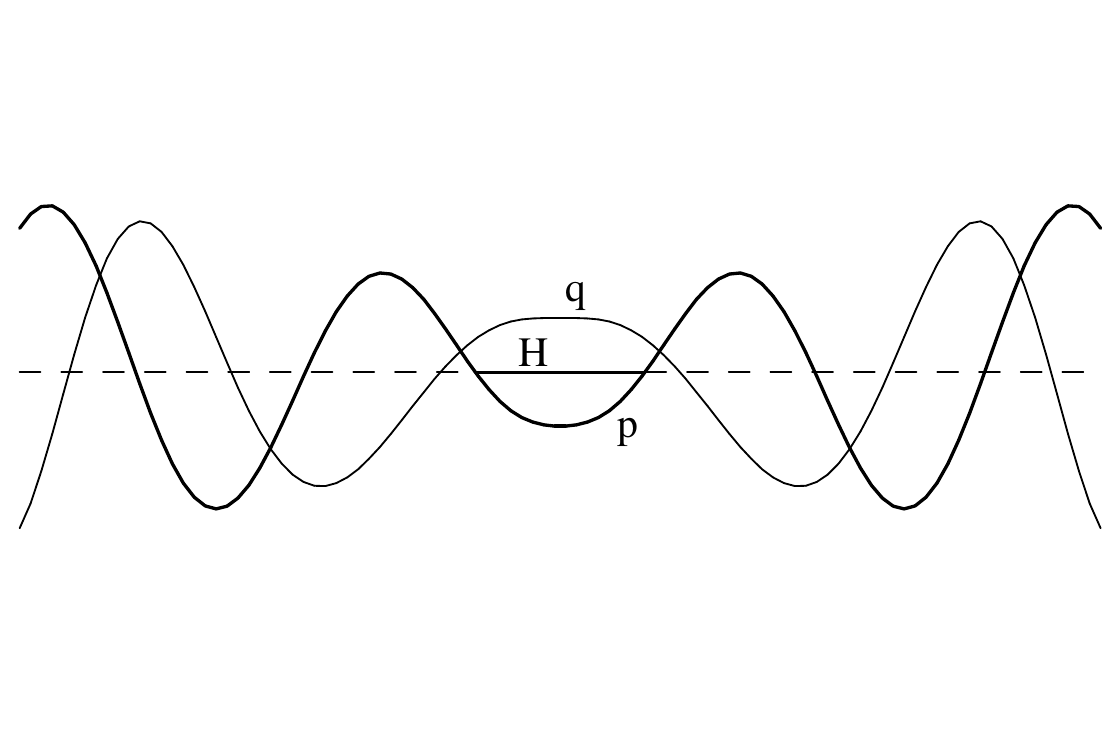}%
}\\
Fig. 4 Polynomial $p$ and a separator $q$%
\end{center}
One can find a separator $q$ by the following method.

\begin{theorem}
\label{Alt}For an arbitrary compact oscillatory set $\mathrm{Z}$ of degree
$m$, there exists\ a polynomial $q$ of degree $<m$ that separates $\mathrm{Z}$
from any hyperbolic point.
\end{theorem}

\textbf{Proof. }The case\textbf{\ }$n=1$ is trivial; we assume that $n>1.$
Take a hyperbolic point $a$ and consider the Euler field $\mathbf{e}_{a}%
=\sum\left(  x_{i}-a_{i}\right)  \partial/\partial x_{i}$ centered at $a.$ The
polynomial $q_{a}\doteq\mathbf{e}_{a}\left(  p\right)  -mp$ has degree $\leq
m-1$ and for any unit $\omega,$
\begin{equation}
q_{a}\left(  a+t\omega\right)  =t^{m+1}\frac{\mathrm{d}}{\mathrm{d}t}\left(
t^{-m}p\left(  a+t\omega\right)  \right)  =tp_{t}^{\prime}\left(
a+t\omega\right)  -mp\left(  a+t\omega\right)  . \label{15}%
\end{equation}
We numerate roots of $p\left(  a+t\omega\right)  $ by $t_{k}=t_{k}\left(
\omega\right)  ,\ k=\pm1,...,\pm\mu$ as in (\ref{16})\ so that $t_{k}$ has the
same sign as $k.$ By (\ref{15}) and Rolle's theorem $q_{a}\left(
a+t\omega\right)  $ has, at least, $m-2$ roots $s_{k}=s_{k}\left(
\omega\right)  ,\ k=\pm1,...,\pm\left(  \mu-1\right)  $ such that
\begin{equation}
t_{k}\leq s_{k}\leq t_{k+1},\ t_{-k-1}\leq s_{-k}\leq t_{-k},\ k=1,...,\mu-1.
\label{13}%
\end{equation}
By continuity it is true for all unit $\omega$ occasionally with non strict
inequalities. Check that there are exactly $m-2$ such zeros $s_{k}$ counting
with multiplicities. For any $\omega\in\mathrm{S}$ except for a set of zero
measure all zeros $t_{k}$ are simple and all inequalities (\ref{13}) are
strict. Suppose that one of the intervals (\ref{13}) say $I_{k}\doteqdot
\left(  t_{k},t_{k+1}\right)  $ contains more than one zero $s_{k}.$ The total
number $r_{k}$ of zeros of $q_{a}$ in $I_{k}$ is odd, since the polynomial
$t^{-m}p\left(  a+t\omega\right)  $ does not vanish in $I_{k}.$ It
follows\ that $r_{k}\geq3$ and $r_{i}\geq1$ for any $i,$ which implies
$\sum_{i}r_{i}\geq m$ . This is not possible since $\mathrm{\deg\ }q_{a}\leq
m-1$. By continuity (\ref{13}) holds also for any unit $\omega$ in the sense
that $q_{a}\left(  a+s\omega\right)  $ has a zero $s$ of multiplicity $m-1,$
if this point is a root of $p\left(  a+s\omega\right)  $ of multiplicity $m$.
It implies that all zeros $s=s_{k}\left(  \omega\right)  ,\ k=1,...,\mu-1$ are
unambiguously defined and by Rouche's theorem are continuous functions of
$\omega.$We have $s_{k}\left(  -\omega\right)  =-s_{-k}\left(  \omega\right)
$ and each variety
\[
\mathrm{W}_{k}=\left\{  x=a+s_{k}\left(  \omega\right)  \omega,\ \omega
\in\mathrm{S}\right\}  ,\ k=1,...,\mu-1
\]
is closed and homeomorphic to a sphere. By (\ref{13}) these varieties separate
hypersurfaces $\mathrm{Z}_{k},k=1,...,\mu$ constructed in Theorem \ref{Con}.
Check that $q\left(  a+t\omega\right)  $ does not vanish for $t\in\left(
t_{-1},t_{1}\right)  .$ We can assume that $p<0$ in $\mathrm{H}$ and have
$q_{a}\left(  a+t_{\pm1}\omega\right)  =t_{\pm}p_{t}^{\prime}\left(
a+t_{\pm1}\omega\right)  \geq0.$ If $q$ vanishes in a point $s\in\left(
t_{-1},t_{1}\right)  $, it must have $\geq2$ zeros, which is impossible since
$\deg q_{a}<m$. This shows that $q_{a}$ separates $p$ from $a.$ It follows
that $q_{a}$ separates $p$ from any other$\ $point $b\in\mathrm{H}$ since the
hyperbolic cavity $\mathrm{H}$ is inside of all ovals $\mathrm{W}%
_{k}.\ \blacktriangleright$

\begin{corollary}
Any separator $q$ of a compact oscillatory set $\mathrm{Z}$ of degree $m=2\mu$
is an oscillatory polynomial with a hyperbolic cavity $\mathrm{G}%
\supset\mathrm{H.}$ The zero set $\mathrm{W}$ of $q$ is the union of
continuous ovals $\mathrm{W}_{1},...,\mathrm{W}_{\mu-1}$ and of a closed
unbounded component $\mathrm{W}_{\mu}$ if $q$ is not strict such that the sets
$\mathrm{H},$ $\mathrm{Z}_{1},\mathrm{W}_{1},\mathrm{Z}_{2},...,\mathrm{Z}%
_{\mu-1},\mathrm{W}_{\mu-1},\mathrm{Z}_{\mu},\mathrm{W}_{\mu}$ are nested (see
figure 1).
\end{corollary}

\textbf{Proof. }Any strict separator has no zeros $x=a+s\omega$ except for
$s=s_{k}$ as in (\ref{13}). If $q$ is not strict it must have exactly one such
real zero, say $s=s_{\mu}\left(  \omega\right)  $, which is defined and
continuous for $\omega\in\mathrm{S}\backslash\mathrm{S}^{\prime},$ where
$\mathrm{S}^{\prime}$ is a subset of dimension $<n-1.$ This function is odd,
since any line $L_{\omega}=\left\{  x=a+s\omega,\ s\in\mathbb{R}\right\}
,\ \omega\in\mathrm{S}\backslash\mathrm{S}^{\prime}$ contains only one such
zero. It follows that $\mathrm{S}^{\prime}$ divides the unit sphere in two
opposite parts $\mathrm{S}_{\pm}$ and $s_{\mu}\left(  \omega\right)
\rightarrow\infty$ as $\omega\rightarrow\mathrm{S}^{\prime}.$ The equation
$x=a+s_{\mu}\left(  \omega\right)  \omega$ defines an unbounded component
$\mathrm{W}_{\mu}$ for $\omega\in\mathrm{S}_{+}.\ \blacktriangleright$

\begin{proposition}
\label{C}Let $p$ be an oscillatory polynomial of degree $m$ such that the
Taylor series of $p\left(  a+y\right)  $ does not contain terms of degree
$m-1$ for a hyperbolic point $a$. Then $q_{a}=\mathbf{e}_{a}\left(  p\right)
-mp$ is a strict separator of $p$. In particular, for any even polynomial $p$
with compact zer0 set the polynomial $q_{0}$ is a strict separator.
\end{proposition}

\textbf{Proof. }The polynomial $q_{a}$ is a strict separator since $\deg
q_{a}\leq m-2$. If $p$ is even and $\mathrm{Z}$ is compact, the unique
hyperbolic$\ $cavity $\mathrm{H}$ is symmetric with respect to the origin and
contains the origin since it is convex. Therefore the second statement follows
from the first one.$\ \blacktriangleright$

Note without proofs few more geometric properties of oscillatory sets.

\begin{proposition}
Any compact oscillatory set $\mathrm{Z}$ that has a strict separator $q$ can
be approximated by regular oscillatory sets $\mathrm{\tilde{Z}}$ whose strict
separators approximate $q.$
\end{proposition}

\begin{center}
\fbox{\includegraphics[
natheight=3.000000in,
natwidth=4.717600in,
height=3in,
width=4.7176in
]%
{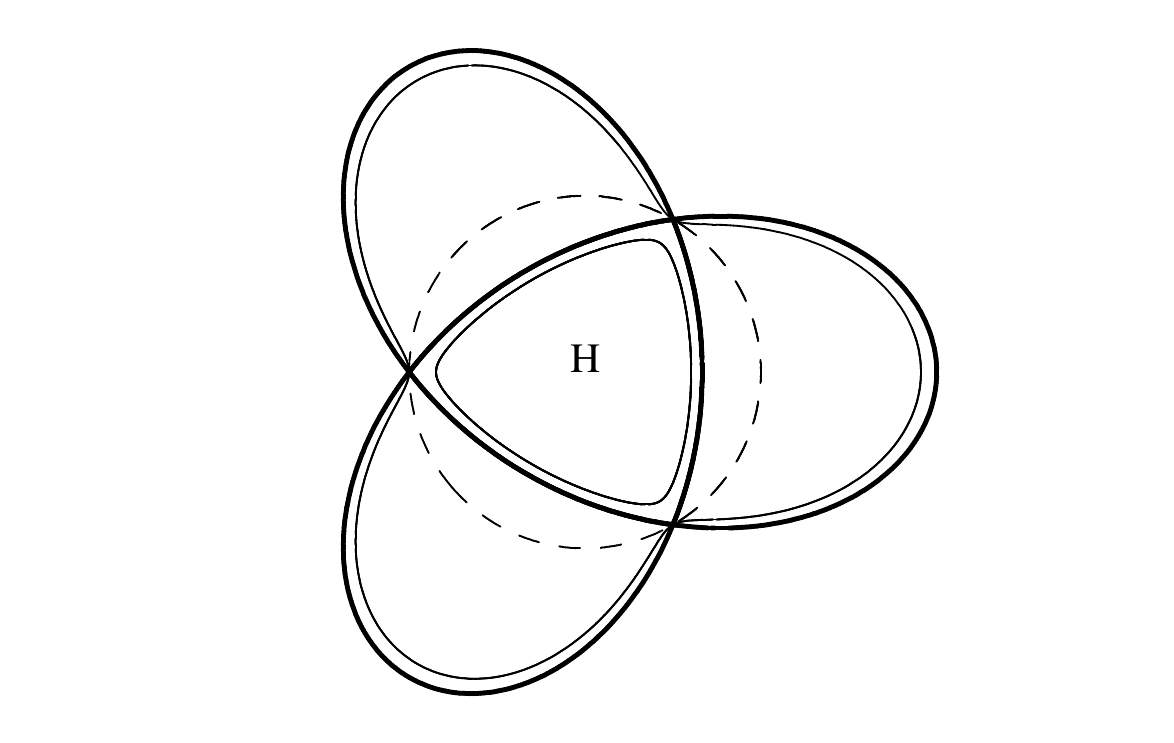}%
}\\
Fig. 5 Hypotrochoid (thick) and a regular approximation (thin)
\end{center}

\textbf{Example 5. }A hypotrochoid given by$\ p\left(  x,y\right)  =4\left(
x^{2}+y^{2}\right)  ^{2}-4x^{3}+12xy^{2}-27\left(  x^{2}+y^{2}\right)  +27$ is
oscillatory and $q=4\left(  x^{2}+y^{2}\right)  -9$ is a strict separator (see
figure 5).

\begin{proposition}
Any compact convex domain $Q$ in $\mathbb{R}^{n}$ can be approximated by
regular hyperbolic$\ $cavities $\mathrm{H}$ of oscillatory sets admitting
strict separators.
\end{proposition}

\section{A generalization of Newton's levitation theorem}

I. Newton \cite{N1} proved that a mass uniformly distributed over a thin
sphere $\mathrm{S}\subset E^{3}$ generates the zero gravitation field inside
the sphere. The same true for the cavity of a solid layer between two
ellipsoids homothetic with respect to the center. P. Dive \cite{D} called a
monoid the layer between any two closed homothetic surfaces with respect to an
interior point. He\ proved that any non-ellipsoidal monoid can not create
levitation in the cavity. V. Arnold \cite{Ar4} constructed a distribution of
electric charge on a regular compact oscillatory set $\mathrm{Z}$ that
generates the zero electrical field in the hyperbolic cavity $\mathrm{H}.$
This distribution, however, has variable sign except if $\mathrm{Z}$ is not an
ellipsoid$\mathrm{.}$ Similar problems for magnetic fields\ were studied in
\cite{VS}. We show that for any oscillatory set $\mathrm{Z}$ that admits a
strict separator, there exists a strictly positive mass distribution on
$\mathrm{Z}$ that generates levitation in $\mathrm{H}$. If $\mathrm{Z}$ is
regular, the set of such mass distributions form a convex cone of dimension $%
\genfrac{\vert}{\vert}{0pt}{}{m-2+n}{n}%
.$ Also layers bounded by close level sets of the corresponding oscillatory
polynomial $p$ generate levitation in a cavity. The key notion is a strict
separator of an oscillatory polynomial.

\begin{theorem}
\label{N}For an arbitrary$\ $oscillatory polynomial $p$ in $E^{3}$ with
compact zero set and any strict separator $q,$ the distribution of mass in
$E^{3}$ with density $\left\vert q\right\vert \delta\left(  p\right)  $
generates the zero gravitation field in the hyperbolic$\ $cavity $\mathrm{H}$
of $p.$
\end{theorem}

\textbf{Proof. }For any $\omega\in\mathrm{S}\backslash\mathrm{S}^{\prime},$
the polynomial $p_{\omega}\left(  t\right)  =p\left(  a+t\omega\right)  $ has
$m$\ real zeros $t_{k}\left(  \omega\right)  ,$ $k=\pm1,\pm2,...,\pm\mu$
numerated as in Theorem \ref{Con}. By the Residue theorem we have%
\begin{equation}
\sum\frac{q\left(  a+t_{k}\omega\right)  }{p_{t}^{\prime}\left(  a+t_{k}%
\omega\right)  }=\sum\mathrm{res}_{t_{k}}\frac{q\left(  a+t\omega\right)
}{p_{t}^{\prime}\left(  a+t\omega\right)  }\mathrm{d}t=-\mathrm{res}_{\infty
}\frac{q}{p}\mathrm{d}t=0. \label{8}%
\end{equation}
According to the numeration a zero $t_{k}$ is positive or negative together
with $k.$ Suppose that $p<0$ and $q>0$ in $\mathrm{H}.$ Signs of
$p_{t}^{\prime}\left(  a+t_{k}\omega\right)  ,k=-\mu,...,-1,1,...,\mu$
alternate and $kp_{t}^{\prime}\left(  a+t_{k}\omega\right)  >0$\ for any odd
$k$ whereas $kp_{t}^{\prime}\left(  a+t_{k}\omega\right)  <0$ with even
$k$.\ The sign of $q\left(  a+t_{k}\omega\right)  $ alternates in the
different way: $q\left(  a+t_{k}\omega\right)  >0$ for odd $k$ and $q\left(
a+t_{k}\omega\right)  <0$ for even $k.$ Therefore we have
\begin{equation}
\frac{q\left(  a+t_{k}\omega\right)  }{p_{t}^{\prime}\left(  a+t_{k}%
\omega\right)  }>0\ \text{for}\ k>0;\text{ }\frac{q\left(  a+t_{k}%
\omega\right)  }{p_{t}^{\prime}\left(  a+t_{k}\omega\right)  }<0\ \text{for}%
\ k<0. \label{17}%
\end{equation}
The sum of these fractions vanishes by (\ref{8}). By Newton's law the
gravitation field decreases at the same rate as a beam of straight rays
diverges. Following Newton's geometrical method we consider a small solid
angle $\mathrm{C}\left(  a,\omega\right)  \subset E$ with vertex at $a$ of
spherical measure $\mathrm{d}\omega.$ The mass of a piece of $\mathrm{Z}%
\cap\mathrm{C}\left(  a,\omega\right)  $ at a point $x=a+t_{k}\omega$ is equal
to
\[
m_{k}\doteqdot\left\vert t_{k}\right\vert ^{2}\frac{\left\vert q\left(
x\right)  \right\vert }{\left\vert \left\langle \omega,\nabla p\left(
x\right)  \right\rangle \right\vert }\mathrm{\Omega}=\left\vert t_{k}%
\right\vert ^{2}\left\vert \frac{q\left(  x\right)  }{p_{t}^{\prime}\left(
x\right)  }\right\vert \mathrm{\Omega.}%
\]
Its contribution to the field at $a\ $is equal to $F_{k}\doteqdot
m_{k}\left\vert t_{k}\right\vert ^{-2}\mathrm{d}\omega$ for $k>0$ and
$F_{k}\doteqdot-\left\vert t_{k}\right\vert ^{-2}m_{k}\mathrm{\Omega}$ for
$k<0,$ since these points are on opposite site of $a.$ By (\ref{17}) the total
of these contributions equals the sum (\ref{8}) times $\mathrm{\Omega},$ hence
cancels. Theorem, follows since this conclusion holds for almost all $\omega$.
$\blacktriangleright$

\textbf{Remark 1. }For the polynomial $p=\left\vert x\right\vert ^{2}-1,$ the
above construction gives $q_{0}=2$. Theorem \ref{N} guarantees levitation in a
sphere generated by a uniform distribution of mass on the sphere. This is
Newton's attraction theorem. If $\mathrm{Z}$ is a compact regular oscillatory
set of degree $m$ and $q$ is a strict separator, then any polynomial
$\tilde{q}$ of degree $m-2$ that is sufficiently close to $q$ is a strict
separator. By Theorem \ref{N} for any such $\tilde{q},$ $\tilde{q}%
\mathrm{d}\xi/\mathrm{d}p$ is a volume form in $\mathrm{Z.}$ Any mass
distribution in $\mathrm{Z}$ that is proportional this form admits levitation
in $\mathrm{H}$.

\textbf{Remark 2. }Theorem \ref{N} is generalized for arbitrary linear space
$\mathbb{R}^{n},$ if the corresponding "gravitation" force generated by a
delta-like mass at the origin has the potential $U=\sigma\left(
\omega\right)  r^{-n+1},$ where $\sigma\left(  \omega\right)  $ is an
arbitrary even function of $\omega\in\mathrm{S}$.

\begin{corollary}
\label{NV}Under conditions of Theorem \ref{N}, if $p<0$ in $\mathrm{H,}$ the
distribution of mass with density $\left\vert q\right\vert \mathrm{d}x$ in the
layer $L\doteqdot\left\{  x:a\leq p\left(  x\right)  \leq b\right\}  $
generates the zero gravitation field in $\mathrm{H}$ for arbitrary $a,b$ such
that $q\neq0$ in $L.$
\end{corollary}

\textbf{Proof}. By Theorem \ref{N}, the density $\left\vert q\right\vert
\delta\left(  p-\lambda\right)  $ generates the zero gravity in $\mathrm{H}$
for any $\lambda$ such that $q$ separates $p-\lambda.$ By Fubini's the same is
true for the density%
\[
\int_{a}^{b}\left\vert q\right\vert \delta\left(  p-\lambda\right)
\mathrm{d}\lambda=\left\vert q\right\vert \mathrm{d}x
\]
supported by the layer $\left\{  a\leq p\leq b\right\}  $ if $q$ separates
$p-\lambda$ for $a\leq\lambda\leq b$ .$\ \blacktriangleright$

A layer generated by a hypotrochoid\ is shown in figure 5.

\textbf{Example 6. }A surface of normals of the system of crystal optics is
given by the equation $p\left(  \xi\right)  =0$ where
\begin{align*}
p\left(  \xi\right)   &  =\left(  \sigma_{1}\xi_{1}^{2}+\sigma_{2}\xi_{2}%
^{2}+\sigma_{3}\xi_{3}^{2}\right)  \left\vert \xi\right\vert ^{2}\\
&  -\left(  \sigma_{3}+\sigma_{2}\right)  \sigma_{1}\xi_{1}^{2}-\left(
\sigma_{1}+\sigma_{3}\right)  \sigma_{2}\xi_{2}^{2}-\left(  \sigma_{1}%
+\sigma_{2}\right)  \sigma_{3}\xi_{3}^{2}+\sigma_{1}\sigma_{2}\sigma_{3}%
\end{align*}
is an even elliptic oscillatory polynomial. The polynomial $q=\mathbf{e}%
_{0}\left(  p\right)  -4p$\ is a strict separator.

\section{Non strict case}

\begin{theorem}
Let $p$ be an elliptic oscillatory polynomial of degree $m$ and $q$ be a
separator. The gravitation field generated by mass distribution in
$\mathrm{Z}$ with the density $\left\vert q\right\vert \delta\left(  p\right)
$ is constant in the hyperbolic cavity and equals%
\begin{equation}
F=-\int_{\mathrm{P}^{n-1}}\frac{q_{m-1}\left(  \omega\right)  }{p_{m}\left(
\omega\right)  }\cdot\omega\ \mathrm{\Omega}, \label{14}%
\end{equation}
where integration van be taken over an arbitrary unit hemisphere.
\end{theorem}

Note that the integrand $q_{m-1}\left(  \omega\right)  /p_{m}\left(
\omega\right)  \cdot\omega$ is an even vector function of $\omega.$

\textbf{Proof.} Let again $\mathrm{C}\left(  a,\omega\right)  $ be a small
solid angle as in Theorem \ref{N}. By (\ref{8}) the gravitation field
generated by points on $\mathrm{Z}\cap\mathrm{C}\left(  a,\omega\right)  $
upon a point $a\in\mathrm{H}$ equals the vector%
\[
\sum_{k=1}^{m}\frac{q\left(  x\right)  }{p_{t}^{\prime}\left(  x\right)
}\cdot\omega\ \mathrm{\Omega}.
\]
where $x=a+t_{k}\omega.$ By the Residue theorem, the sum of scalars is equal
to%
\[
\sum_{k=1}^{m}\frac{q\left(  x\right)  }{p_{t}^{\prime}\left(  x\right)
}=-\mathrm{res}_{\infty}\frac{q}{p}\mathrm{d}t=-\frac{q_{m-1}\left(
\omega\right)  }{p_{m}\left(  \omega\right)  }.
\]
Integrating over any unit hemisphere, we get (\ref{14}). $\blacktriangleright$

\end{document}